%% file: faroprep.tex
\documentstyle[sprocl,epsfig,amstex]{article}

\input{latexuseful209.tex}

\input{wspc-refstyle.tex}
\def\kr{{\kern -0.1em}}

\begin{document}

\input{faro-title.tex}





\section{Introduction}

One of the long standing problems in multiparticle dynamics is
the description of multiplicity distributions and correlation
functions within a common formalism which can emphasize
the dynamical processes underlying multiparticle production.
In this talk, I will illustrate one step in this direction,
by discussing how the sign oscillations of the ratio
of factorial cumulant moments to factorial moments of
the Multiplicity Distribution (MD)
can be related to the dynamics of the early stage of
partonic evolution in \ee\ annihilation.

In Section \ref{sec:definitions} the relevant observables
are defined and illustrated with examples; in Section
\ref{sec:qcd} and \ref{sec:experiments} theoretical 
calculations and experimental results are reviewed;
in Section \ref{sec:common}
the shoulder structure in the MD
and the oscillations of moments are related
to hard gluon radiation; 
finally in Section \ref{sec:2jets}
the effect of flavour quantum numbers on
the moments of the MD in 2-jet events is discussed;
conclusions are drawn at the end.

\section{Definitions and examples: the effect of truncation}
\label{sec:definitions}

In this section the definitions of some
differential and integral observables and their relationships
will be recalled; see reference~\cite{DeWolf:rep} for further details.
The definitions, summarized in Table \ref{tab:defs},
concern $n$-particle distributions:
the {\em exclusive} ones, $P_n(y_1,\ldots,y_n)$
where {\em all} $n$ particles produced at 
rapidities $y_1,\ldots,y_n$ are observed,
and the {\em inclusive} ones, $Q_n(y_1,\ldots,y_n)$
where {\em at least} $n$ particles are observed.
These quantities are of course related:
\begin{multline}
 P_n(y_1,\ldots,y_n) = Q_n(y_1,\ldots,y_n) +\\ 
   \sum_{m=1}^{\infty}
	\frac{(-1)^m}{m!} \int dy_1'\ldots dy_m' Q_{n+m} (y_1,\ldots,y_n,
	y_1',\ldots,y_m')				\label{eq:lvh}
\end{multline}
From the inclusive distributions, via cluster expansion, one gets
the {\em correlation functions} $C_n(y_1,\ldots,y_n)$,
thus subtracting from $Q_n$ the statistical, uninteresting correlations
due to combinations of lower order distributions $Q_{n-1}\ldots Q_1$;
however there remain correlations related to fluctuations in the
number of particles $n$:\,\cite{correl} it is therefore only
in an approximate sense that one can say that when $C_2$ is positive
particles like to cluster together (which one relates to the dynamics
of the process) and when $C_2$ is negative particles
like to stay away from each other (which one relates to the effect
of conservation laws).
\begin{table}[t]      
\caption[Definitions]{Definitions of relevant differential
(on the left) and integral (on the right) quantities; $\sigma$
is the inclusive cross section, $\sigma_{\mathrm inel}$ the
inelastic cross section, $y_i$ is the rapidity of the $i$-th particle.
The integrals in
the right-hand column are over the full phase space.}
\label{tab:defs}
\def\somespace{\omit\vphantom{x}&\omit\cr}
\small
\begin{tabular}{ll}
\somespace
\hline
\somespace
Exclusive distribution  &  Multiplicity distribution\cr
$\displaystyle P_n(y_1,\ldots,y_n)$ &
$\displaystyle P_n = \frac{1}{n!} \int dy_1\ldots dy_n P_n(y_1,\ldots,y_n)$\cr
\somespace
\hline
\somespace
Inclusive distribution  &  Factorial moments\cr
$\displaystyle Q_n(y_1,\ldots,y_n) = \frac{1}{\sigma_{\mathrm inel}} 
  \frac{d^n\,\sigma}{dy_1\ldots dy_n}$ &
$\displaystyle F_n = \int dy_1\ldots dy_n Q_n(y_1,\ldots,y_n)	$\cr
\somespace
\hline
\somespace
Correlation function, {\it e.g.}  &  Factorial cumulant moments\cr
$\displaystyle C_2(y_1,y_2) = Q_2(y_1,y_2) - Q_1(y_1) Q_1(y_2)$ &
$\displaystyle K_n = \int dy_1\ldots dy_n C_n(y_1,\ldots,y_n)	$\cr
\somespace
\hline
\end{tabular}
\end{table}

\begin{figure}[!p]    
\mbox{\epsfig{file=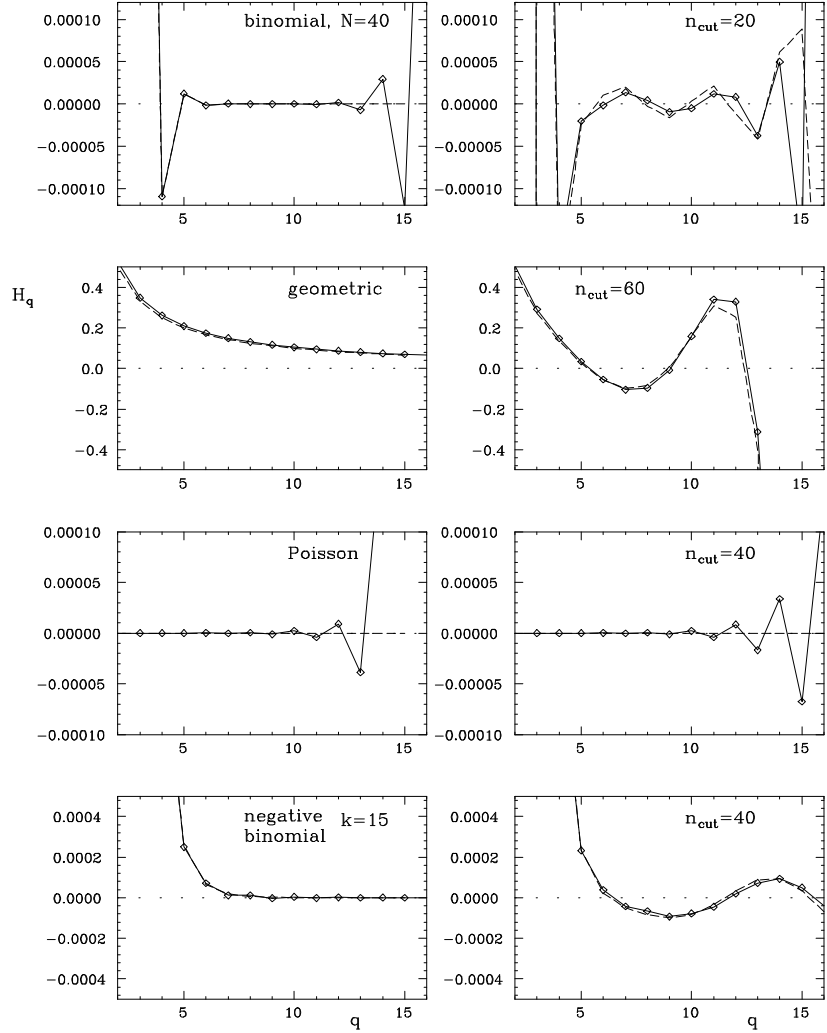,bbllx=53pt,bblly=72pt,%
bburx=531pt,bbury=678pt,height=13.9cm}}
\caption[Examples of moments]{The ratio $H_q$ for the 
most common discrete distributions.
All the distributions are
chosen to have the same average number of particles (namely, $\nbar = 10$);
other parameters, if any, are listed in the figure.
The plots in the left column correspond to the full distributions 
(dashed lines), and to the even-only distributions (solid lines joining
the diamonds). The plots in the right column correspond to the same
distribution after truncation has been taken into account
as in Eq.~\ref{eq:truncation}; the effect of the truncation
changes with the highest multiplicity $\ncut$:\,\cite{hqlett} 
the values chosen here are such that the discarded part is
always less than 1\% of the cumulative distribution 
function.}\label{fig:moments}
\end{figure}

By integrating the differential observables just discussed one
obtains the integral observables which will be the subject of
the rest of this talk (see Table \ref{tab:defs}):
the {\em multiplicity distribution},
the {\em factorial moments}
and the {\em factorial cumulant moments}.
They are linked by the following relationships,
analogous to Eq.~\ref{eq:lvh}, which allow to obtain
the moments directly from the MD:
\begin{gather}
  F_n = \sum_{r=n}^{\infty} r(r-1)\cdots(r-n+1) P_r \\
  K_n = F_n - \sum_{r=1}^{n-1} {n-1 \choose r} K_{n-r} F_r  \label{eq:Kn}
\end{gather}
It is apparent from the definitions that $F_n$ and $K_n$
receive contributions from events
with at least $n$ particles, which means that 
moments of high order
are very sensitive to the tail of the MD. In particular,
as will be shown in the following, 
it is interesting in this respect to study the behaviour of the ratio
\begin{equation}
  H_q \equiv \frac{K_q}{F_q}				\label{eq:Hq}
\end{equation}
as a function of the order $q$:
it is qualitatively different for different distributions and
turns out to be suitable for analytical calculations (see Section
\ref{sec:qcd}). In order to illustrate these properties,
Figure \ref{fig:moments} (left column, dashed lines) shows
the ratio $H_q$ for
some of the most common discrete distributions.
It is worth pointing out that for the Poisson distribution,
which one usually associates with a lack of dynamical correlations, 
the ratio $H_q$ is zero for $q > 1$; 
for the geometric and for the negative binomial distributions,
which one usually associates with dynamical effects because
they give rise to positive correlations,
the ratio $H_q$ is always positive, but decreases toward zero; for
the binomial distribution, which one usually associates with 
conservation laws because of negative correlations, the ratio $H_q$
changes sign according to the parity of $q$.
Before applying these considerations to the data in full phase space,
one should remember that only even multiplicities are allowed for
charged particles. In order to show how the suppression of the
odd multiplicities affects the moments,
the left column in Figure \ref{fig:moments}
presents also the ratio $H_q$ vs the order $q$
subject
to the condition that $P_n = 0$ if $n$ is odd (solid lines and points):
it is seen
that there is a small distortion with respect to the values of the
full distributions.

As already mentioned, the ratio $H_q$ probes the tail
of the MD. Unfortunately, the tail is the
most difficult part to measure experimentally, because only a finite
number of events can be collected. This results in a MD which is
truncated at some point:
\begin{equation}
  \tilde P_n \propto \begin{cases} 
     P_n & \text{if $n \leq \ncut$} \\
     0  & \text{otherwise} 
    \end{cases}					\label{eq:truncation}
\end{equation}
where the proportionality factor ensures proper normalization.
The moments of high order are affected: the factorial
moments are smaller than in the full distribution, and one finds that
the factorial cumulant moments oscillate in sign as the order
increases.\cite{hqlett} An example of this effect can be
seen again in 
Figure \ref{fig:moments};
its importance for experimental MD's
will be discussed in Section \ref{sec:experiments}.

\section{The ratio $H_q$ in perturbative QCD}\label{sec:qcd}

The first suggestion that the ratio $H_q$ could be useful in
the study of MD's came from analytical calculations in 
perturbative QCD,\cite{Nechitailo,Dremin} where
it was shown that 
one expects oscillations of the ratio $H_q$ as a function of the order $q$.
In the following I will summarize the derivation in the framework of pure
gluodynamics,\cite{Nechitailo} starting from the evolution
equations for the generating function
$\cG(\cY;z) = \sum_{n=0}^{\infty}(1+z)^n P_n(\cY)$:
\begin{multline}
  \frac{d \cG(\cY;z)}{d \cY} = \int_0^1  d x \left( \frac{1}{x} -
 	(1-x)[2-x(1-x)] \right)
	\left( \frac{2 \NC \alpha_s(\cY)}{\pi} \right) 
	 \times \\
      \big[\cG(\cY+\ln x;z) \cG(\cY+\ln(1-x);z) - \cG(\cY;z) \big]
						\label{eq:evolution}
\end{multline}
where the first term in parentheses is the 
DGLAP kernel for the process $g\to gg$ with the emitted gluon carrying
a fraction $x$ of the parent momentum, $\cY = \ln(\kt/Q_0)$ 
is the evolution variable ($\kt$ the jet's transverse momentum and
$Q_0$ the cutoff, which limit the integration interval via
$x(1-x)\kt > Q_0$),
$\alpha_s(\cY)$ is the running coupling constant and
$\NC$ is the number of colours. Notice that the terms in the square
brackets take recoil into account, thus going beyond the pure
Double Log Approximation (DLA).
In order to solve Eq.~\ref{eq:evolution} for the moments of the distribution
one expands it in powers of $\cY$, remembering that,
when calculating correlations of order $q$,
one should use $q\gamma$ as expansion parameter (because the NLO
contribution to such correlations is proportional to $q\gamma$,
rather than $\gamma$; $\gamma$ is here the anomalous dimension).
Assuming asymptotic KNO scaling, so that
only the average number of partons depends on $\cY$, while
the normalized moments of higher order are constant as $\cY\to\infty$,
one calculates now the derivatives in $z=0$ of
the generating function $\cG$, obtaining the factorial moments; 
the factorial cumulant 
moments are then computed via Eq.~\ref{eq:Kn}. The normalization
of these moments is not fixed but can be eliminated by taking
their ratio. One finds for the ratio $H_q$ 
a negative minimum at $q \approx 5$ and
sign oscillations at larger $q$. This result has been
qualitatively improved by including
an estimate of the contribution of the vertices $q\to qg$ and
$g\to q\bar q$;\,\cite{Dremin2} further confirmation comes
from the exact solution of the full evolution equations including
both quarks and gluons in the case of fixed coupling.\cite{DreminHwa2}
It could be noted that the oscillations found in this last case
happen around the (monotonically decreasing) value of $H_q$
of a negative binomial distribution (NBD) with $k \approx 5$.
On the other hand, it is interesting to point out that in
DLA the oscillations disappear and the ratio $H_q$ is very close
to that of a NBD with $k \approx 3$.

However, before doing comparison of partonic results
with experimental data, one should investigate what the
possible role of hadronization is. Here I will just recall~\kr\cite{VietriRU}
that if the simplest possible Generalization to Local Parton
Hadron Duality~\kr\cite{AGLVH:2} (GLPHD), which requires the proportionality
of all inclusive distributions at parton (p) and hadron (h) level
via a single parameter $\rho$:
\begin{equation}
  Q_n^{\mathrm h}(y_1,\ldots,y_n) = \rho^n Q_n^{\mathrm p}(y_1,\ldots,y_n),
\end{equation}
is used, it is easy to show, by direct substitution into 
the formulae of Table \ref{tab:defs},
that one obtains:
\begin{equation}
  H_n^{\mathrm h} = \frac{\rho^n K_n^{\mathrm p}}{\rho^n F_n^{\mathrm p}}
  	= H_n^{\mathrm p}
\end{equation}
Thus, if the GLPHD hadronization prescription is used, the parton
level result on the ratio $H_q$ can be directly applied to the
hadrons: in this case, it is seen that the QCD prediction
for \ee\ annihilation~\kr\cite{Dremin2} fails quantitatively.

\section{The ratio $H_q$ in experiments}\label{sec:experiments}

The theoretical work described in the previous Section triggered
the analysis (a posteriori) of available data on MD's in order
to extract the features of the ratio $H_q$.\cite{Gianini}
\begin{figure}[t]         
\begin{center}
\begin{minipage}{7cm}
\mbox{\epsfig{file=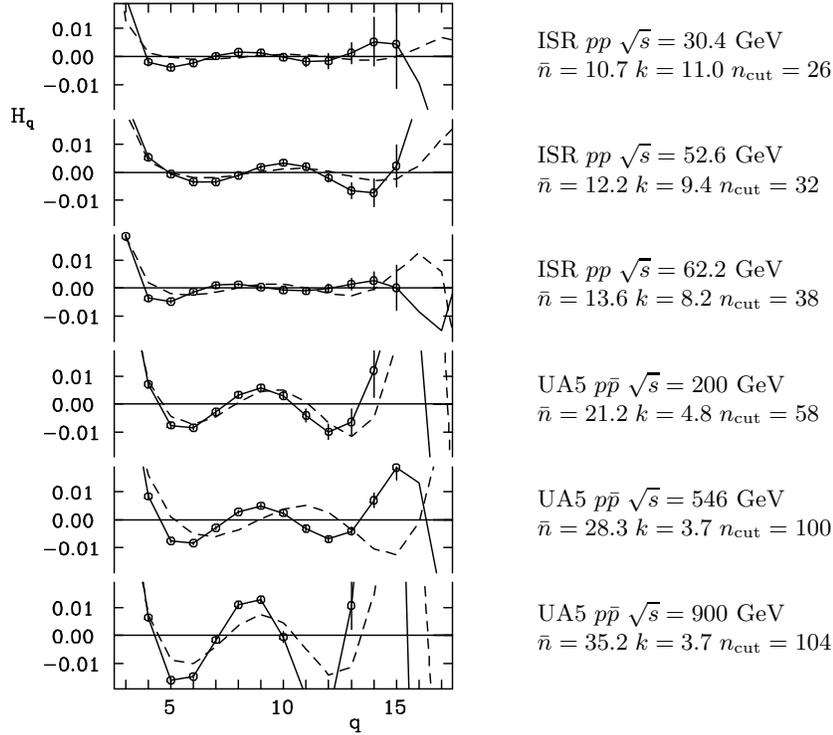,bbllx=140pt,bblly=96pt,bburx=392pt,%
bbury=512pt,height=10cm}}
\end{minipage} 
\begin{minipage}{4cm}\small %
ISR $pp$ \roots{30.4}\\ $\nbar=10.7 \; k=11.0 \; \ncut=26$\\[0.75cm]
ISR $pp$ \roots{52.6}\\ $\nbar=12.2 \; k= 9.4 \; \ncut=32$\\[0.75cm]
ISR $pp$ \roots{62.2}\\ $\nbar=13.6 \; k= 8.2 \; \ncut=38$\\[0.75cm]
UA5 $p\bar p$ \roots{200}\\ $\nbar=21.2 \; k= 4.8 \; \ncut=58$\\[0.75cm]
UA5 $p\bar p$ \roots{546}\\ $\nbar=28.3 \; k= 3.7 \; \ncut=100$\\[0.75cm]
UA5 $p\bar p$ \roots{900}\\ $\nbar=35.2 \; k= 3.7 \; \ncut=104$\\
\vspace{0.3cm}
\end{minipage}\end{center}
\caption[Hq moments in hadron-hadron collisions]{The ratio $H_q$
vs the order $q$ for a collection of experiments in hadron-hadron
collisions.\cite{hhdata} The solid line joins data points and is
drawn to guide the eye only. The dashed line shows the prediction
of the NBD which fits well the MD's, after the truncation effect
has been taken into account.}\label{fig:gianini}
\end{figure}
It should be immediately said that this is not an easy task:
in order to measure the MD in a modern experiment an `unfolding'
procedure has to be used;\cite{DEL:1} the resulting
published MD has correlations between adjacent bins which are
not taken into account when moments are extracted from it
without knowledge of the correlation matrix. This makes the
calculation of the errors on the moments thus obtained
very difficult: it should be considered as an order
of magnitude estimate rather than a precise determination.
Keeping this in mind, one can proceed to review some
experimental results. The main point here is that
oscillations of very different amplitudes are seen
in the data in all reactions, in most cases (but not all!
see below) compatible with being due to the truncation of the 
MD. As an illustration,
in Figure \ref{fig:gianini} some results relative to hadron-hadron
collisions are shown: large oscillations are found.
The MD's from which these moments are extracted are well
described (with the exception of UA5 data at \roots{900})
by a NBD. Also shown in the same figure
is the ratio $H_q$ predicted by the fitted NBD
after taking into account the effect of truncation
(as in Eq.~\ref{eq:truncation}): it is seen that the oscillations
thus obtained are of the same order of magnitude as the data,
so that no further dynamical effects, beyond those described by a 
single NBD, are apparent (see also the analysis in~\cite{Biyajima:pp}).

The case is different in \ee\ annihilation data at the $Z^0$ peak,
as exemplified in Figure \ref{fig:sld}, where data from the
SLD Collaboration~\kr\cite{SLD} are compared 
with the predictions of a NBD
(dotted line) and a truncated NBD (dot-dashed line). It should be 
mentioned that in this plot the errors on the data are the
statistical errors as published by the SLD Collaboration,
which take into full account all correlations from the unfolding
matrix. Clearly the truncated distribution cannot describe the data:
it will be shown in the next section how one can relate
these oscillations to hard gluon radiation.

\section{Common origin of the shoulder structure and
of the oscillations}\label{sec:common}

A very interesting feature in the data on 
charged particles MD's at the $Z^0$ peak
is the shoulder structure which is clearly
visible in the intermediate multiplicity range.\cite{DEL:2,OPAL,ALEPH}
This peculiar behaviour appears evident if one looks at the
residuals (difference between the data and fitted values, in units
of standard deviations) with respect to a NBD (or even a Lognormal
Distribution):\,\cite{SLD} the curve starts below the data, goes above
and then below again. A pattern is seen, instead of a random
sign and size of residuals.
\begin{figure}[t]     
\begin{center}
\mbox{\epsfig{file=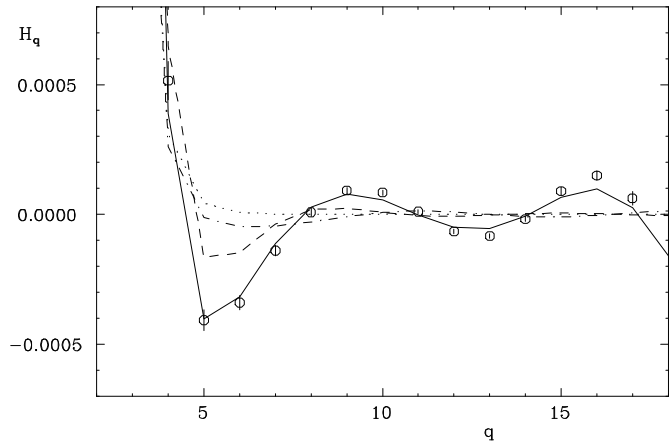,bbllx=112pt,bblly=288pt,bburx=512pt,%
bbury=534pt,width=9cm}}
\end{center}
\caption[Hq moments by SLD]{
The ratio of factorial cumulant moments over factorial moments,
$H_q$ as a function of $q$;
experimental data (diamonds) from the SLD 
Collaboration~\kr\cite{SLD}
are compared with the predictions of several parameterizations, 
with parameters fitted to the data on MD's: a full NBD (dotted line); 
a truncated NBD (dot-dashed line); 
sum of two full NBD's as per Eq.~\ref{eq:formula} (dashed line); 
sum of two truncated NBD's as per Eq.~\ref{eq:truncformula} (solid line).
}\label{fig:sld}
\end{figure}

The DELPHI Collaboration has shown~\kr\cite{DEL:4} that
the shoulder structure in the MD in $e^+e^-$ annihilation can be explained
by  the superposition of the MD's coming from
events with 2, 3 and 4 jets, as identified by a suitable jet-finding
algorithm, and that the MD's in these classes of events are well
reproduced by a single NBD for a range of values of the jet-finder
parameter, $\ymin$.  It is thus suggested that the shoulder is
associated with the radiation of hard gluons resulting in the
appearance of one of more extra jets in the hadronic final state.  One
should also recall that a shoulder structure similar to the one
observed in $e^+e^-$ annihilation has been observed in $p\bar p$
collisions at high energies~\kr\cite{UA5:3,Rimondi} and was shown to
be well described by a 5-parameter parametrization in terms of the
weighted superposition of two NBD's.\cite{Fug}

Following these observations, 
a parametrization of the MD which is the weighted sum
of two components, one to be associated with 2-jet events and
one to be associated with events with 3 or more jets,
was proposed.\cite{hqlett:2} The weight
in this superposition is then the fraction of 2-jet events,
which is experimentally determined,  not a fitted parameter.
This decomposition depends of course on the definition of jet, and
in particular
on the value of the parameter which controls the jet-finding algorithm. The
DELPHI Collaboration~\kr\cite{DEL:4} 
has used the JADE algorithm and published
values for the 2-jet fraction and for the MD's at $\ymin$ = 0.02, 0.04,
0.06, 0.08.

As for the particular form of the MD in the two components, 
the NBD was chosen because it has successfully been 
fitted to the data for the samples of events with fixed number of 
jets.\cite{DEL:4} 
In practice a fit was performed to the MD's with a four parameter
formula: 
\begin{equation}
  P_n \propto \begin{cases} \alpha P_n^{\NB}(\nbar_1,k_1) + (1-\alpha)
	P_n^{\NB}(\nbar_2,k_2) & \text{if $n$ is even}\\
  0 & \text{otherwise}\end{cases}			\label{eq:formula}
\end{equation}
Here $P_n^{\NB}(\nbar,k)$ is the standard NBD of parameters $\nbar$ and $k$;
notice that the charge conservation law is taken into account.
The
proportionality factor is fixed by requiring the proper normalization 
for $P_n$.

Results of the fit to the data of four experiments can be summarized
as follows:
the $\chi^2$ per degree of freedom are equal to or smaller than 1, 
and values of the parameters are consistent between
different experiments; 
they are also consistent with those obtained
by the Delphi Collaboration 
in fitting their 2-jet and 3-jet data separately with
a NBD.

Once the fits to the MD's have been done,
one can compare the experimental data on $H_q$'s
with the values obtained from 
the fitted MD's by using a formula that takes  into account the
truncation effect too:
\begin{equation}
  \tilde P_n \propto \begin{cases} P_n & \text{if 
        ($n_{\mathrm{min}} \le n \le n_{\mathrm{max}}$)}\\
  0 & \text{otherwise}\end{cases}		\label{eq:truncformula}  
\end{equation}
where $n_{\mathrm{min}}$ and $n_{\mathrm{max}}$ are the minimum and
maximum observed multiplicity, and a proportionality factor ensures
proper normalization. A very good agreement with the data is obtained,
as exemplified in Figure \ref{fig:sld} (solid line).
Notice that it is not possible to
reproduce the behaviour of the ratio $H_q$ without taking into account the
limits of the range of the  available data:
this can be seen again
in Figure~\ref{fig:sld}, where the dashed line corresponds to 
Eq.~\ref{eq:formula} and does not agree with the data.

It can thus be concluded that the observed behavior 
of $H_q$'s results from the
convolution of two different effects, a statistical one, i.e., the
truncation of the tail due to the finite statistics of data samples, and 
a physical one, i.e., the
superposition of two components.
The two components can be related to 2- and 3-jet
events, i.e., to the emission of hard gluon radiation in the early stages of
the perturbative evolution.
Notice that as the energy increases, the number of components also
grows so that, asymptotically, the oscillations should be
washed out, in agreement with the DLA expectations.

\section{Effects of flavour in 2-jet events}\label{sec:2jets}

A simple check of the picture described in the previous section
consists in looking at the behaviour of the ratio $H_q$ 
in 2-jet events, using the MD given by the DELPHI 
Collaboration:\,\cite{DEL:4} one does not expect oscillations.
It is found~\kr\cite{Gianini} that oscillations are present,
although their amplitude is one order of magnitude smaller
than in the full sample. 
\begin{table}[b]     
\caption[Fits to 2-jet events]{Parameters and $\chi^2$/NDF of 
the fit to experimental data on 2-jet events
MD's from the DELPHI Collaboration~\kr\cite{DEL:4} with
the weighted superposition  of two NBD's with the 
same parameter $k$ (Eq.~\ref{eq:2nbd});
the weight used is the fraction of $b\bar b$ events. 
Results are shown for different values of  the jet-finder parameter $\ymin$.} 
\label{tab:2jet-fits}
\newcommand\pp{$\pm$}
 \begin{center}
 \begin{tabular}{||c|c|c|c||}
\hline
              &  $\ymin$ = 0.01  & $\ymin$ = 0.02  & $\ymin$ = 0.04 \\ 
\hline
$\bar n_l$    & 16.81\pp 0.21 & 17.22\pp 0.15 & 17.98\pp 0.15 \\ 
$\bar n_b$    & 20.26\pp 1.71 & 21.96\pp 1.57 & 23.61\pp 1.64 \\ 
$k$           & 124\pp 51     & 145\pp 53     & 120\pp 33 \\
$\chi^2$/NDF  & 17.4/16       & 12.6/16       &  27.5/20  \\
$\delta_{bl}$ & 3.44\pp 0.83  & 4.6\pp 0.5    & 5.6\pp 0.5 \\ 
\hline
 \end{tabular}
 \end{center}
\end{table}
If one looks at the MD itself in 2-jet events,
one discovers that the residuals, with respect to a NBD, show a structure
similar to the one seen in the full sample;
furthermore, the oscillations
in the ratio $H_q$ cannot be described by truncating one NBD.\cite{hqlett:3}
The hint that there could
be a substructure comes once again from  DELPHI data:\,\cite{DEL:bb} 
they showed that the MD in a single hemisphere in a sample
enriched in $e^+e^- \to b\bar b$ events is identical in shape to the MD
in a sample without flavour selection, except for a shift
of 1 unit:
\begin{equation}
  P_n^{(b)} = P_{n-1}^{\text{({\it all})}}
\end{equation}
This effect could be due to weak decays of  B-hadrons.\cite{Jorge}
This result hints at a possible substructure in terms of events
with heavy quarks and events with light quarks,
each sample contributing to the MD with one NBD.
One can therefore try a parametrization of the form~\kr\cite{hqlett:3}
\begin{equation}
  P_n(\bar n_l, \bar n_b, k) = 
      \alpha_b P_n^{\mathrm{NB}}(\bar n_b, k) + 
      (1 - \alpha_b ) P_n^{\mathrm{NB}}(\bar n_l, k)
						\label{eq:2nbd}
\end{equation}
where now $\alpha_b$ is the fraction of $b\bar b$ events at the
$Z^0$ peak, measured at LEP to be approximately 0.22. Notice
that, as suggested by the DELPHI result, the parameter $k$
is the same in the two NBD's: this is therefore a fit with
three parameters. 
The parameters of the fits and the corresponding $\chi^2$/NDF  
are given in Table~\ref{tab:2jet-fits}
for different  values of the jet resolution parameter $\ymin$;
the fit has been performed taking the charge conservation
law into account similarly to Eq.~\ref{eq:formula}.
A really accurate description of experimental data is achieved. 
Notice that the best-fit value for the difference between the average
multiplicities in the two samples, $\delta_{bl}$, 
also given in Table~\ref{tab:2jet-fits}, is quite large. This difference
grows with increasing $\ymin$, i.e., with increasing contamination
of 3-jet events. By looking at the residuals one concludes that
the proposed parametrization can reproduce  the experimental data on MD's 
very well, as no structure is visible.\cite{hqlett:3} 
Furthermore this parametrization
describes well also the ratio $H_q$.\cite{hqlett:3}
It is also remarkable that only the average number of particles
depends on flavour quantum numbers, whereas the NBD parameter $k$ is
flavour-independent.

One can thus conclude that the examination of the 
behaviour of the ratio $H_q$ as a function of $q$
has allowed to link the final hadronic level with the flavour
composition of the event.

\section{Conclusions}\label{sec:conclusions}

The ratio of factorial cumulant moments to factorial moments,
$H_q$, is a good observable for exploring substructures
in hadronic final states. It is possible to calculate its
behaviour in perturbative QCD and, after making allowance
for the effect of finite statistics in the data, to extract
dynamical information about the first stages of the
perturbative evolution. In fact, hard gluon radiation
explains the substructures observed in the full sample of
events in \ee\ annihilation at the $Z^0$ peak; flavour
dependent properties explain the 
additional substructures observed
in the sample of 2-jet events. It is also remarkable
that in this analysis the most elementary substructures
are well described by negative binomial distributions
(down to 2-jet events of fixed flavour). A final point
should be made, as the fits discussed in this talk have been
performed on the published data, and therefore could not
take into account the full correlations: the hope is that
the interesting results thus obtained will
spawn experimental work on the original data.

\section*{Acknowledgments}
I would like to thank the organizers of this meeting for
creating a
warm and fruitful atmosphere, and Jorge Dias de Deus
in particular
for inviting me to give this talk. I would
also like to thank Alberto Giovannini and Sergio Lupia
for their critical reading of this manuscript.

\section*{References}
\input{faro.ref}

\end{document}

%% file: latexuseful209.tex
\def\ee{$e^+e^-$}               

\newcommand{\roots}[1]{$\sqrt{s} = #1$ GeV} 

\def\nbar{\bar n}               

\def\pt{p\kern -.2pt\lower 4pt\hbox{\tiny T}}    

\def\p0{P_0(\Delta y)}

\def\ymin{y_{\mathrm{min}}}
\def\NB{{\mathrm{NBD}}}

\def\kt{k_{\perp}}
\def\ncut{n_{\mathrm cut}}

\def\NF{{\mathcal N}_{\kern -1.9pt f}}
\def\NC{{\mathcal N}_{\kern -1.7pt c}}

\def\cY{{\mathcal Y}}
\def\cG{{\mathcal G}}

{ \newcount\hour \newcount\minutes \newcount\hourint
  \minutes=\time \hour=\time \divide\hour by 60
  \hourint=\hour   \multiply\hourint by -60  \advance\minutes by \hourint
  \xdef\timestamp{\the\hour:\ifnum\minutes>9\else0\fi\the\minutes}
}  


%% file: wspc-refstyle.tex
\def\journal#1#2#3#4#5{{\it #1}\if-#2{}\else #2 \fi{\bf #3}, #5 (#4)}

\def\bookproceedings#1#2#3#4#5{{\it #1} (#2) #3: #4\if-#5{}\else, p.~#5\fi}

\def\preprint#1#2#3{``#1'', \if-#2{}\else #2 \fi preprint #3}
\def\proceedings#1#2#3#4#5{Proceedings of the \bookproceedings{#1}{#2}{#3}{#4}{#5}}

%% file: faro-title.tex
\thispagestyle{empty}
\begin{center}
\hfill LU TP 96-27\\
\hfill October 20th, 1996\\
\vspace{2.4cm plus 1.6cm minus 0.5cm}

{\large STRUCTURES IN MULTIPLICITY DISTRIBUTIONS AND OSCILLATIONS OF MOMENTS}
\vspace{0.8cm plus 1.0cm}

        R. Ugoccioni \footnote[1]{E-mail: roberto@@thep.lu.se}\\
{\it Department of Theoretical Physics, University of Lund,\\
   S\"olvegatan 14A, S-22362 Lund, Sweden}
\vspace{2cm plus 0.8cm minus 0.5cm}

ABSTRACT\\
\end{center}

\begin{quote}\noindent The possibility to relate
multiplicity distributions and their moments, as measured
in the hadronic final state in $e^+e^-$ annihilation,
to features of the initial partonic state
is analyzed from a theoretical and phenomenological point
of view. Recent developments on the subject are discussed.
\end{quote}

\vspace{2.5cm plus 1.2cm minus 0.8cm}
\begin{center}
Invited talk presented at the\\
 XXVI International Symposium on Multiparticle Dynamics\\ 
(Faro, Portugal, September 1-5, 1996)
\end{center}

\newpage